\documentclass{article} 
\usepackage{arxiv_style}
\usepackage{times}
\usepackage{float}
\usepackage{xcolor}
\usepackage{graphicx}
\usepackage{graphics}
\usepackage{url,hyperref}
\usepackage[numbers]{natbib}
\usepackage{physics}
\usepackage{qcircuit}
\usepackage{authblk}
\usepackage{amsthm}
\usepackage{amsmath}
\usepackage{amssymb}
\usepackage{multicol}
\usepackage{multirow}
\usepackage{subfigure}

\renewcommand{\vec}[1]{\mathbf{#1}}
\newtheorem{theorem}{Theorem}
\theoremstyle{definition}

\theoremstyle{remark}

\begin{document}

\title{Variational quantum algorithms on cat qubits}
\author[1]{Anne-Solène Bornens}
\author[1,\thanks{michel.nowak@thalesgroup.com}]{Michel Nowak}

\affil[1]{\textit{Thales Research and Technology, Palaiseau, France}}

\maketitle
\thispagestyle{empty}
\begin{abstract}
Variational Quantum Algorithms (VQA) have emerged with a wide variety of applications.
One question to ask is either they can efficiently be implemented and executed on existing architectures.
Current hardware suffers from uncontrolled noise that can alter the expected results of one calculation.
The nature of this noise is different from one technology to another.
In this work, we chose to investigate a technology that is intrinsically resilient to bit-flips: cat qubits.
To this end, we implement two noise models.
The first one is hardware-agnostic -- in the sense that it is used in the literature to cover different hardware types.
The second one is specific to cat qubits.
We perform simulations on two types of problems that can be formulated with VQAs (Quantum Approximate Optimization Algorithm (QAOA) and the Variational Quantum Linear Solver (VQLS)), study the impact of noise on the evolution of the cost function and extract noise level thresholds from which a noise-resilient regime can be considered.
By tackling compilation issues, we discuss the need of implementing hardware-specific noise models as hardware-agnostic ones can lead to misleading conclusions regarding the regime of noise that is acceptable for an algorithm to run.
\end{abstract}
\begin{multicols}{2}
\section{Introduction}
\label{sec:introduction}
Quantum computing offers two types of algorithms. The ones that solely rely
on the preparation and control of quantum states -- referred to as end-to-end quantum
algorithms -- and those relying on a classical feedback loop -- referred to as
Variational Quantum Algorithms (VQA).
Although end-to-end algorithms come with proven theoretical speedups assuming the availability of enough computational resources, they are hardly implementable on existing hardware.
During the current hardware development phase (Noisy Intermediate Scale Quantum phase~\cite{preskill2018quantum}), operations performed on physical systems will suffer from uncontrolled behavior, which can be considered as noise.

The impact of noise on VQAs has already been studied with a hardware-agnostic noise model, where Pauli gates are introduced before and after the application of unitaries~\cite{wang2021noise}.
However, because hardware architectures significantly differ in the way information is encoded, transformed, and finally measured, we can infer that the impact of noise on the performance of an algorithm is also hardware specific.
To support this statement, we can cite~\cite{henriet2020robustness}, which already studies the impact of some physical properties of a device on the performance of a VQA. We thus take the same research path, which intends to tighten the gap between the fields of algorithm design and the physics at play in the hardware architectures.

Among the different hardware proposals for quantum computation, cat qubits~\cite{guillaud2022quantum} provide with an interesting feature: when executing a set of identified gates, the information encoded in the cat states remains protected from bit-flip errors.
The goal of the present work is to investigate this specificity by looking at the convergence of VQAs.
In order to reach this goal, we first implement noise models that represent as faithfully as possible the physics of cat qubits.
We then formulate a set of two problems of interest (optimization and linear system inversion), associated with cost functions. We finally conclude on the impact of noise on their minimization on a large number of configurations. This was made possible by the utilization of the large computing cluster at hand.

We structure the work in the following way: Section~\ref{sec:cat_qubits} first recalls the necessary elements to understand how cat qubits are resilient to a particular type of errors (bit-flips) for a given set of gates. Section~\ref{sec:noise_model} regroups the noise models and provides with the characterization of the Hadamard gate which remains unstudied in the literature. By completing the set of noise models, we are finally able to study the impact of uncontrolled operations on VQAs in Section~\ref{sec:numerical_experiments}. We conclude in Section~\ref{sec:conclusion}.

\section{Quantum computing with cat qubits}
\label{sec:cat_qubits}
Quantum computing architectures with the discrete variable formalism (qubits) are usually built on physical systems that have two levels of energy.
These two levels represent the ground and the excited state of an artificial atom (in the case of superconducting qubits) or a real one (in the case of trapped cold atoms).
However, it has been shown that more complex states can be chosen to encode information.
In particular, information can be encoded in the coherent states of a harmonic oscillator~\cite{mirrahimi2014dynamically}.
This work exclusively focuses on this technology.

\subsection{Encoding information}
Let $\ket{\alpha}$ denote a coherent state of complex amplitude $\alpha$.
We follow the notations from~\cite{guillaud2019repetition}, where the convention for the cat state is taken so that:
\begin{equation}
    \ket{0} \approx \ket{\alpha}
\end{equation}
and
\begin{equation}
    \ket{1} \approx \ket{-\alpha}
\end{equation}
The idea of encoding $\ket{0}$ and $\ket{1}$ in coherent states is a first step towards designing a bit-flip-resilient qubit.

\subsection{Protecting information}
The second idea to fully protect information from jumping from $\ket{0}$ to $\ket{1}$ spontaneously is to couple the system with an other one stabilizing it.
In the following, we will thus adopt the notions from Chapter 4 of~\cite{cohen2012processus} where a physical system described by the harmonic oscillator (called the storage) is coupled to a larger system called the reservoir.
In the case of cat qubits, the reservoir is designed so as to impose specific conditions to the storage. These conditions are made so that the storage gains or loses photons by pairs. From the expression of coherent states, the information encoded in $\ket{0}$ will be protected from transforming to $\ket{1}$ if the rate at which photons are gained or lost by pair dominates the rate at which errors of other type happen.

\begin{figure*}[ht]
    \subfigure[Absorption of two photons by the storage\label{four_wave_absorption}]{
        \includegraphics[width=\linewidth]{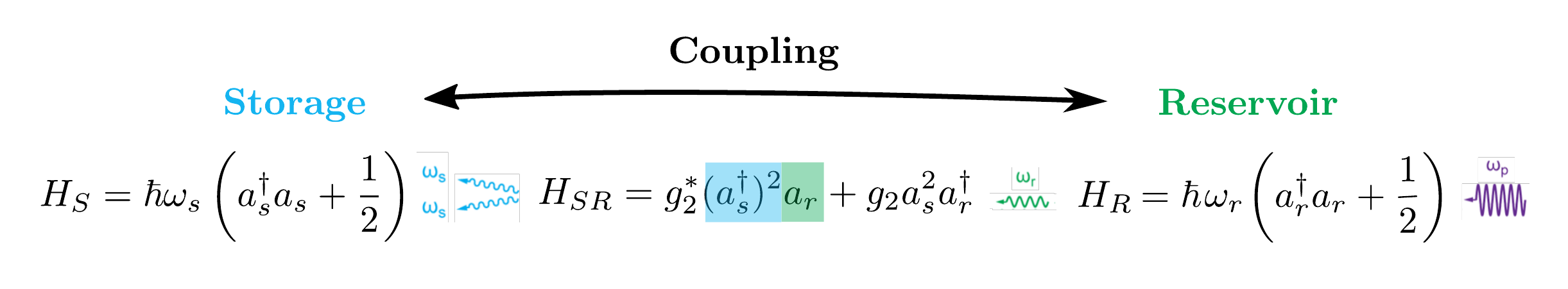}
    }
    \subfigure[Emission of two photons by the storage\label{four_wave_emission}]{
        \includegraphics[width=\linewidth]{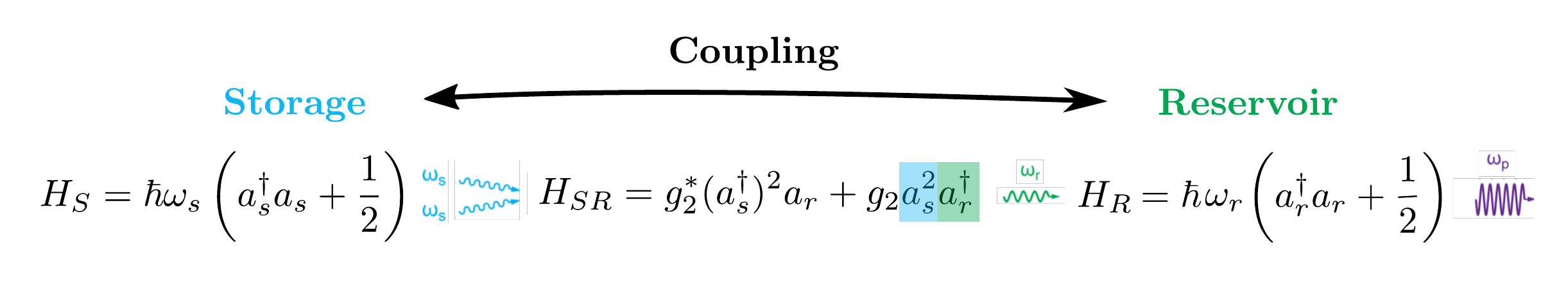}
    }
    \caption{Conversion process between the storage and the reservoir~\cite{leghtas2015confining}}
    \label{fig:four_wave_process}
\end{figure*}

Let us write the hamiltonians for the entire system.
Let $\mathcal{S}$ be a storage system.
Consider that it is coupled to a reservoir $\mathcal{R}$. The hamiltonian of the total system writes:
\begin{equation}
    H = H_S + H_R + H_{SR}
\end{equation}
where the storage hamiltonian is:
\begin{equation}
    H_S = \hbar \omega_{s} \left(a_s^\dag a_s + \frac{1}{2}\right)
\end{equation}
and the reservoir hamiltonian follows the same expression:
\begin{equation}
    H_R = \hbar \omega_{r} \left(a_r^\dag a_r + \frac{1}{2}\right)
\end{equation}
with $\omega_s$ and $\omega_r$ respectively the associated frequencies of the storage and the reservoir.
$H_{SR}$ -- the interaction hamiltonian -- can be designed such as it takes the form derived in Chapter 4 of \cite{lescanne2020engineering}:
\begin{equation}
    H_{SR} = g_2^* (a_s^\dag)^2 a_r +g_2 a_s^2 a_r^\dag
\end{equation}
where $g_2$ is a coupling constant.
$H_{SR}$ involves the square of the annihilation and the creation operators of the storage, which guarantees the desired two photon gain or loss condition.
We illustrate the interactions in Figure~\ref{fig:four_wave_process} by following the same notations of reference~\cite{leghtas2015confining}, which introduces a method to engineer these conditions in practice.
The reservoir is imposed with a radiation of frequency $\omega_p$ such that:
\begin{equation}%
    \omega_p = 2\omega_s - \omega_r%
\end{equation}%
Figure~\ref{four_wave_absorption} illustrates how photons are gained in pairs by the storage via the coupling with the reservoir, while Figure~\ref{four_wave_emission} shows how photons are emitted by the storage.
Mathematically explaining how this process leads to an intrinsic protection against bit-flips is extensively covered in~\cite{guillaud2022quantum}, whereas experimental data can be found in~\cite{lescanne2020exponential}.

\subsection{Applying gates}
This intrinsic noise bias of cat qubits needs to be conserved even during the execution of gates. Such gates that maintain this noise bias are called bias-preserving gates ~\cite{guillaud2022quantum}. The idea of cat qubits is thus to exclusively use this kind of gates so as to maintain a protection against the bit-flips along the computation.
Although the set of bias-preserving gates is large, it does not contain gates such that the Hadamard gate, or rotations of an arbitrary angle around the $X$ axis (except for the X-rotation of angle $\pi$, which is bias-preserving).
In such cases, indeed, bit-flip errors will be introduced via conversion of the phase-flip type errors into bit-flip errors through the depolarizing gates.

The idea behind cat qubits is to design logical gates~\cite{guillaud2019repetition} from these bias preserving gates, by introducing error correction codes designed to reduce the phase-flip errors.
In this work, we make the assumption that gates are only applied at the physical level.
We thus do not resort to error correction techniques, so as to study the possibility to compute solutions in an intermediate regime, where the number of available qubits is not large enough to allow for error correction.

\begin{figure}[H]
    \centering
    \includegraphics[width=0.8\linewidth]{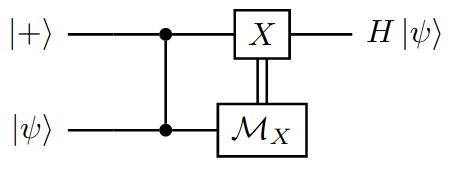}
    \caption{Hadamard circuit on cat qubits architectures adapted from~\cite{guillaud2019repetition}. The initial state is $\ket{\psi}$.}
    \label{fig:hadamard_on_cat_qubits}
\end{figure}

In order to perform some operations on variational circuits, we need the Hadamard gate.
We are aware that introducing a depolarizing gate will remove one part (if not all) of the advantage of cat qubits, but we need it to compile a VQA. 
Consider the circuit depicted in Figure~\ref{fig:hadamard_on_cat_qubits}.
It shows how to perform a Hadamard gate on cat qubits. Although the original paper in which this circuit is proposed stands for logical gates, one can still use it at the physical level.
An ancilla register is prepared in the $\ket{+}$ state.
Then, a $CZ$ gate follows between the ancilla register and the register on which we wish to apply the Hadamrd gate.
An $X$ gate is finally applied on the ancilla register conditioned on the measurement in the $X$ basis of the initial state register.
The ancilla register finally contains the desired output state.

\section{Noise models}
\label{sec:noise_model}

We now tackle the problem of implementing a relevant noise model for our study of VQAs.
Section~\ref{sec:cat_qubits_noise_model} regroups the main contributions of this work, where the implementation of a noise model specific to cat qubits is provided.
Section~\ref{sec:arbitrary_noise_model} then covers a theoretical and hardware-agnostic noise model proposed in~\cite{wang2021noise} along with some adaptations to better compare with the cat qubits model.

\subsection{Cat qubits model}
\label{sec:cat_qubits_noise_model}
Let us now introduce the noise model for the cat qubits technology.
As mentioned in Section~\ref{sec:cat_qubits}, the encoding of cat qubits relies on the fine manipulation of coherent states constrained to gain or lose photons by pairs.
One first source of error originates from non respecting this constraint, thus losing photons one by one. 
A second source of error comes from the fact that the gates are applied within a certain time-frame, leading to so-called non adiabatic effects. The gate duration is optimized to maximise the gate fidelity taking into account these two phenomena~\cite{guillaud2019repetition}.
Note that errors will happen asymmetrically on the control or target qubits, and this asymmetry is taken into account in the probabilistic noise models of~\cite{guillaud2021error}.

\begin{figure}[H]
    \centering
    \includegraphics[width=0.8\linewidth]{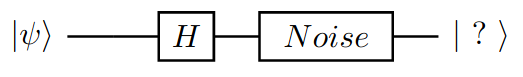}
    \caption{Equivalence of noisy hadamard circuit on cat qubits for simulation}
    \label{fig:hadamard_on_cat_qubits_with_errors}
\end{figure}

For the Hadamard gate, we consider the circuit in Figure~\ref{fig:hadamard_on_cat_qubits} proposed in \cite{guillaud2019repetition} at the logical level. We wish to use this strategy to implement the Hadamard gate at the physical level.
We want a simple realistic error model of this Hadamard gate for our simulations. This model could be directly derived from the circuit. However, this would require to add at least one ancilla to each circuit we simulate. To overcome this burden, we choose to seek for a single qubit noise model, as presented in Figure~\ref{fig:hadamard_on_cat_qubits_with_errors}.
As in~\cite{guillaud2021error} we only consider the errors at the first order in the sense that we neglect the contributions of multiple errors accumulating in the small Hadamard circuit. This gives a first approximation of the level of noise that one can expect from this circuit.
The corresponding model is presented in bold in Table~\ref{tab:final_noise_model}.

\begin{table*}[t]
    \resizebox{\textwidth}{!}{%
    \begin{tabular}{|c||cc|cc|cc|cc|cc|cc|cc|cc|}
    \hline
    \multirow{8}{*}{Physical level} & \multicolumn{2}{c}{$I$} & \multicolumn{2}{|c}{$\mathcal{P}_{\ket{+}}$} & \multicolumn{2}{|c}{$Z(\theta)$} & \multicolumn{2}{|c}{$X$} & \multicolumn{2}{|c|}{$\mathbf{H}$} & \multicolumn{2}{|c}{$CZ$} & \multicolumn{2}{|c}{CNOT} & \multicolumn{2}{|c|}{Toffoli} \\
    \cline{2-17}
      & Error & Probability & Error & Probability & Error & Probability & Error & Probability & \textbf{Error} & \textbf{Probability} & Error & Probability & Error & Probability & Error & Probability \\
    \cline{2-17}
     & $I$ & $1-p$ & $I$ & $1-p$ & $I$ & $1-p$ & $I$ & $1-p$ & $\mathbf{I}$ & $\mathbf{1-5p}$ & $I$ & $1-2p$ & $I$ & $1-4p$ & $I$ & $1-6p$ \\
     & $Z$ & $p$ & $Z$ & $p$ & $Z$ & $p$ & Z &  $p$ & $\mathbf{Z}$ & $\mathbf{3p} $ & $Z_1$ & $p$ & $Z_1$ & $3p$ & $Z_1$  & $p$ \\
     & & & & & & & & & $\mathbf{X}$ & $ \mathbf{2p} $ & $Z_2$ & $p$ & $Z_2$ & $p/2$ & $Z_2$ & $p$ \\
     & & & & & & & & & & & & & $Z_1Z_2$ & $p/2$ & $Z_3$ & $p/2$ \\
     & & & & & & & & & & & & & & & $CZ_{12}$ & $3p$ \\
     & & & & & & & & & & & & & & & $CZ_{12}Z_3$ & $p/2$ \\
    \hline
    \end{tabular}}
    \caption{Final error model. The bold probabilities correspond to what one needs for VQAs. The explicit expression of most of the probabilities can be found in \cite{guillaud2021error} and \cite{chamberland2022}.}
    \label{tab:final_noise_model}
\end{table*}
\subsection{Hardware-agnostic model}%
\label{sec:arbitrary_noise_model}%
Recent work makes the assumption that noise can be modelled by adding Pauli gates between each unitary in the circuit~\cite{wang2021noise}.
They assume that each unitary is taken as being one layer of the circuit.
Even if the nature of the layer depends on the algorithm (QAOA~\cite{farhi2014quantum}, VQLS~\cite{bravo2019variational}, VQC~\cite{schuld2020circuit}, ...), we can say that it encapsulates one application of a chosen ansatz.
We will call this model the hardware-agnostic layer-wise model. One modification that we apply compared to ~\cite{wang2021noise} is that we restrict the errors to be of $X$ and $Z$ type only. This is to be able to compare the noise models more appropriately.

This model allowed to derive potential predictions on the performance of the gradient descent in variational algorithms.
We here recall the main result of this work.
Consider the probabilities of adding a $X$, $Y$ of $Z$ on each qubit after the application of a layer of the VQA: $q_X, q_Y, q_Z$.
Define $p$ as being the maximum of this set of probabilities.
\begin{equation}
    p = \sqrt{\max\{q_X,q_Y,q_Z\}}
\end{equation}
Given these assumptions, an upper bound on the partial derivative of the cost function has been derived and can indicate at which regime the algorithm will suffer from poor convergence.
\begin{theorem}[Upper bound on the partial derivative~\cite{wang2021noise}]
\label{th:upper_bound}
Consider an L-layered ansatz. Suppose that local Pauli noise with parameter p acts before and after each layer with parameter $\theta$. Then the following bound holds for the partial derivative of the noisy cost function.
\begin{equation}
    \left|\frac{\partial\mathcal{C}}{\partial \theta}\right| \leq A\,  n^{1/2}p^{L+1}
    \label{eq:nibp}
\end{equation}
where A is a constant of the problem that does not depend on the number of qubits $n$ and on the noise level $p$.
\end{theorem}

The cat qubits specific model adds noise at the application of each gate.
This makes the agnostic model a priori more optimistic than the specific one.
We thus decide to apply the noisy gates in the hardware-agnostic model at each gate so that to be able to compare to the cat qubits specific model. We will refer to this noise model as the hardware-agnostic gate-based noise model. 

Finally, because hardware providers do not propose the same set of native gates, we chose to constrain the hardware-agnostic model to be compiled with the same gates as the cat qubits one. This model will be adapted if the Toffoli gate is present in the algorithm, and separate simulations with the gate based model will be carried out with and without allowing the Toffoli gate to be present in the set of basis gates for the compilation step since this gate is native with cat qubits but not with most of other existing technologies.

All operations that can be performed in parallel are done in parallel and not sequentially.

\subsection{Comparing the models}
With these models in hand, comparing the cost function convergence will allow to extract the impact of the asymetric noise of cat qubits compared to a uniformly depolarizing channel.

We need to set the definition of the noise level $p$ which will vary from $10^{-9}$ to $10^{-1}$. We chose to set it to the phase flip probability of the $Z$ gate, which is what is taken in~\cite{guillaud2021error} to introduce the noise model. We justify this choice by stating that this unit of error is the unique thing that is common between the different noise models.

One last consideration that needs to be tackled is that the cat qubits model is designed to take into account errors on idle qubits while other experience an application of a gate.
In the cat qubits model, a phase flip probability is applied with probability $p$.
We chose to apply the same scheme to the hardware-agnostic model, but with a depolarizing probability of $p$ for the phase flip error and $p$ for the bit-flip error.

We will conclude on the relevance of this choice at the moment of the results' analysis.

\section{Numerical Experiments}
\label{sec:numerical_experiments}
The idea of VQAs is to encode the problem of interest into the minimization of a cost function with some part of the circuit being dependent on varying parameters~\cite{cerezo2021variational}. For each application, the structure of the circuit is chosen so that the expectation of some measurement is proportional to the cost function. Depending on the outcome of the experiment, a classical optimizer can tune the parameters according to the gradient of the cost function.

Without being comprehensive, but with the hope to cover the widest range of applications, we choose to focus on two particularly challenging setups.
First, optimization problems are frequently being taken to analyse VQAs~\cite{wang2021noise, henriet2020robustness}, we thus include the Quantum Approximate Optimization Algorithm (QAOA)~\cite{farhi2014quantum} into our study.
Then, a more recent VQA has emerged to solve linear systems~\cite{bravo2019variational}: the Variational Quantum Linear Solver (VQLS) which can be helpful to solve computational physics problems.

For each parameterized algorithm, we select setups where we can increase the size of the problem so that to exhibit the limits of convergence.

\subsection{Noise models implementation}
In order to perform the simulations with the two noise models, we need to implement them.
On a quantum computer, the noise would arise at each sample of the circuit.
The correct strategy to simulate the noise according to the failure rate probability $p$ would thus be to randomly introduce gates with respect to this probability at each shot. This can be performed with the custom noise models proposed by Qiskit.

In order not to miss certain gates that do not have a noise model implemented, for example $R_y$ gates, we transpile the circuit with the following gates: $R_z$, $CX$, $CCX$, $X$, $Z$, $H$. When the $CCX$ gate will be present, we will also remove it from the set of available gates for transpilation to test for the potential appearance of discrepancies compared to the native toffoli gate in the cat qubits technology.

\subsection{Quantum Approximate Optimization Algorithm}
We first focus on the MaxCut problem, that can be solved with QAOA~\cite{qiskit_qaoa}.
Let $G$ be an unweighted graph on $n$ vertices. A cut is a partition of the graph's vertices into two complementary sets $\mathcal{S}$ and $\mathcal{T}$. The MaxCut problem consists in finding a cut such that the number of edges between $\mathcal{S}$ and $\mathcal{T}$ is as large as possible.

\subsubsection{Algorithm implementation}
\begin{figure}[H]
    \centering
    \includegraphics[width=\linewidth]{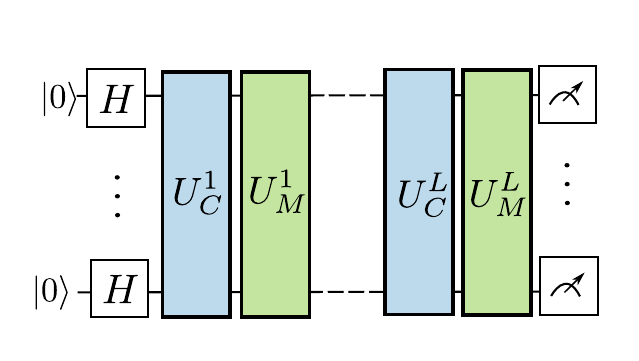}
    \caption{Circuit for QAOA}
    \label{fig:qaoa_circuit}
\end{figure}

We wish to formulate this problem so that it can be solved with QAOA.
Figure~\ref{fig:qaoa_circuit} recalls the structure of the circuit used to solve an optimization problem. Alternating layers of two successive unitaries are applied to a register initially prepared in the $\ket{+}^{\otimes n}$ state.
The first unitary is the mixing one and is built on the following hamiltonian by introducing degrees of freedom $\beta$:
\begin{equation}
U_M(\beta) = e^{-i\beta H_M} \; \; \; \text{with} \; \; \; \beta\in [0, 2\pi]
\end{equation}
and
\begin{equation}
    H_M = \sum_{i=1}^n X^i
\end{equation}
The second unitary is the problem one and is built so that to encode the problem of interest.
It will also be parameterized by another set of degrees of freedom $\gamma$:
\begin{equation}
U_C(\gamma) = e^{-i\gamma H_C} \; \; \; \text{with} \; \; \; \gamma \in [0, 2\pi]
\end{equation}
In order to find an expression for $H_C$, we need to derive a cost function that will encode the MaxCut problem when minimizing it.
To do so, we define a classical cost function $\mathcal{C}(x)$
\begin{equation}
    \mathcal{C}(x) = 1 -\frac{1}{2 n_E(G)}\sum_{(i,j)\in E(G)} (1 - x_i x_j)
    \label{eq:cost_maxcut}
\end{equation}
where $x=\{x_i\}_{i \in [\![1,n]\!]}$ is the ensemble of the vertices, such as $x_i = 1$ if $ x_i \in \mathcal{S}$ and $x_i=-1$ if $x_i \in \mathcal{T}$. $E(G)$ is the set of edges of the graph and $n_E(G)$ is the number of edges of $G$. The normalization by $n_E(G)$ is added to allow comparisons between different graphs of the same number of nodes.

In order to map this problem to QAOA, we define the problem hamiltonian as:
\begin{equation}
H_C = 1 -\frac{1}{2 n_E(G)}\sum_{(i, j) \in E(G)} (1 - Z^i Z^j)
\end{equation}
with $Z^i$ the $Z$ operator that acts on the $i$th qubit.
Alternately applying the two unitaries with $L$ repetitions, we get the parameterised final state of the problem:
\begin{equation}
|\psi_L(\vec{\gamma},\vec{\beta})\rangle =  U_M^L(\beta_L)U_C^L(\gamma_L)\cdots U_M^1(\beta_1)U_C^1(\gamma_1) \ket{+}^{\otimes n}
\end{equation}
We finally measure all qubits and compute the cost function:
\begin{equation}
    \mathcal{C}_\text{QAOA}(\gamma, \beta) = \bra{\psi_L(\vec{\gamma},\vec{\beta})
    }
    H_C
    \ket{
        \psi_L(\vec{\gamma},\vec{\beta})
    }
\end{equation}

\subsubsection{Results analysis}
The idea is to observe how the introduction of noise into the quantum circuit of the algorithm has an impact on its optimized solution and to
analyse in theses conditions the influence of the size of the graph ($n \in [4;9]$), the number of layers ($L \in [1;10]$) and the noise strength ($p \in [10^{-9};10^{-1}]$). For the sake of simplicity, we fix the number of qubits to 5 in the main text.

We execute different series of simulations by varying these parameters.
In order to make sure we do not work on particular graphs, each configuration is averaged on 100 randomly generated graphs of the same size, connecting each pair of vertices with probability $0.6$.
We can then observe the final cost function value after convergence in different setups.

\begin{figure*}[t]
    \centering
    \subfigure[Layer-wise hardware-agnostic\label{fig:maxcut_qaoa_hardware_agnostic_layer_wise}]{\includegraphics[width=0.32\linewidth]{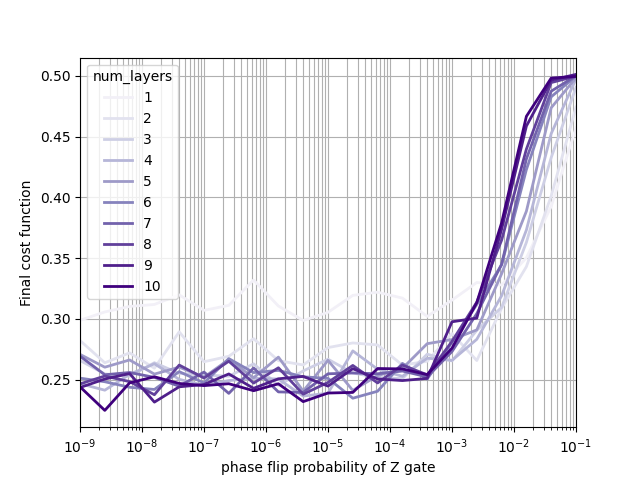}}
    \subfigure[Parallel gate-based hardware-agnostic\label{fig:maxcut_qaoa_hardware_agnostic_parallel}]{\includegraphics[width=0.32\linewidth]{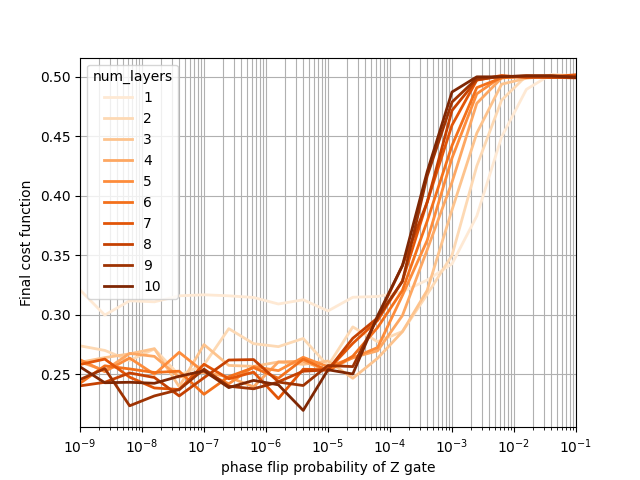}}
    \subfigure[Parallel gate-based cat qubits\label{fig:maxcut_qaoa_cat_qubits}]{\includegraphics[width=0.32\linewidth]{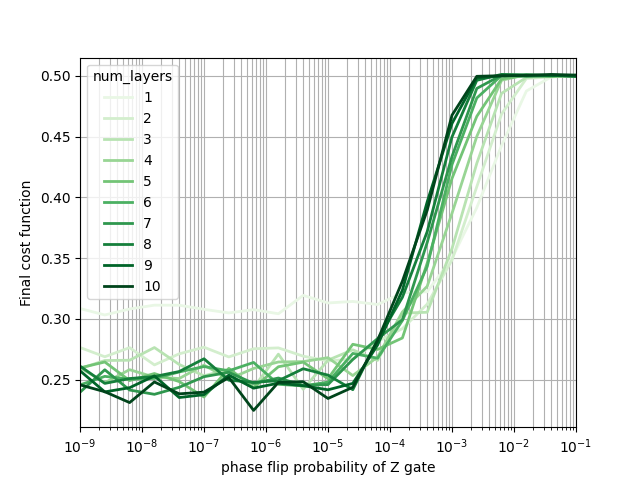}}\\
    \caption{MaxCut QAOA: Evolution of the final cost function reached after convergence depending on the number of layers ranging from 1 to 10 for different noise models ~\ref{fig:maxcut_qaoa_hardware_agnostic_layer_wise} layer-wise hardware-agnostic. ~\ref{fig:maxcut_qaoa_hardware_agnostic_parallel} parallel hardware-agnostic. ~\ref{fig:maxcut_qaoa_cat_qubits} parallel cat qubits model.}
    \label{fig:maxcut_qaoa}
\end{figure*}
We start by looking at the layer-wise hardware agnostic model which is the current reference in the literature~\cite{wang2021noise}.

In Figure~\ref{fig:maxcut_qaoa_hardware_agnostic_layer_wise}, we fix the number of qubits to 5 and vary the number of layers starting from 1 to 10. We observe the final value of the cost function obtained after convergence against the level of noise introduced in the simulation.
For large levels of noise, we identify a regime in which the noise is strong enough so that the cost function converges to $0.5$ which corresponds to a random choice of edges cut.
Depending on the number of layers, the final value of the cost function will start diverging from $0.5$.
It then enters a transient regime where the cost function exponentially decreases towards a reference level - which we will refer to as the noise resilient regime.

Figure~\ref{fig:maxcut_qaoa_hardware_agnostic_parallel} shows the same results for the hardware-agnostic gate-based model. The trend of the curves is the same, however, the thresholds are shifted towards lower values. This is due to the fact that noise is introduced with the same probability at each gate instead of each layer with the same noise model (uniformly depolarizing). We draw a first conclusion on this result: the noise model of~\cite{wang2021noise} is relevant to obtain the trend of the results even in an approximated regime where noise is introduced layer-wise.

Figure~\ref{fig:maxcut_qaoa_cat_qubits} shows the results for the cat qubits model. We observe that the trend is the same as the ones for the hardware-agnostic models.

\begin{figure}[H]
    \centering
    \subfigure[5 qubits]{\includegraphics[width=0.98\linewidth]{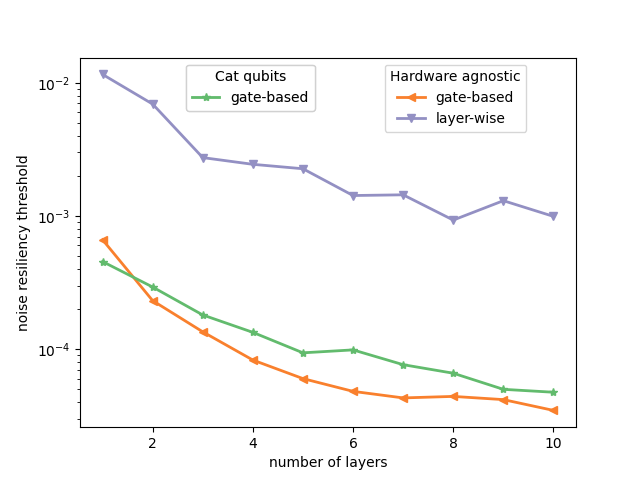}}
    \caption{MaxCut QAOA noise resiliency thresholds evolution against the number of layers for 5 qubits}
    \label{fig:qaoa_noise_resiliency_thresholds_5_qubits}
\end{figure}
We can then extract noise resiliency thresholds from which we can consider the algorithm to be unaffected by the presence of noise.
We thus implement a threshold detection routine that extracts the value from which the convergence enters the transient regime.

Figure~\ref{fig:qaoa_noise_resiliency_thresholds_5_qubits} shows the evolution of the threshold against the number of layers for the two hardware agnostic models and for the cat qubits model. Appendix~\ref{sec:qaoa_noise_resiliency_thresholds} regroups all the configurations simulated.
We observe that the thresholds shifts to lower values as the number of layers increases.
This could be explained by the Theorem~\ref{th:upper_bound}, which stands that the performance of the algorithm might degrade exponentially with the number of layers considering noise and $\left|\partial C/\partial\theta\right|\leq D p^{L+1}$ with $D \in \mathbb{R}$.
We note that the hardware-agnostic model compiled in parallel (gate-based) and the cat qubits model overlap at low values of number of layers but starts diverging as the number of layers increases.
The thresholds for the cat qubits model are slightly higher that the thresholds of the hardware-agnostic one. Appendix~\ref{sec:qaoa_noise_resiliency_thresholds} confirms this observation when varying the number of qubits up to 10.
The layer-wise model reaches the noise resilient regime earlier than the others, and this was expected given that noise is introduced at each layer and not at each time step of the circuit.

\subsection{Variational Quantum Linear Solver}
We wish to study an algorithm that outputs an estimation $\hat{x}$ of the solution to the problem
\begin{equation}
    Ax = b
\end{equation}
where $A$ is a matrix and $b$ a target vector.

\subsubsection{Algorithm implementation}
\begin{figure}[H]
    \centering
    \includegraphics[width=0.8\linewidth]{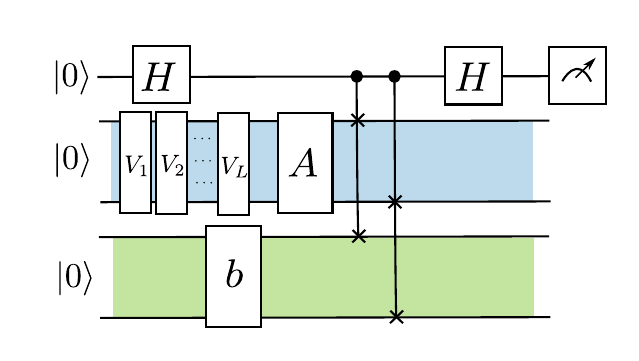}
    \caption{Circuit for VQLS~\cite{bravo2019variational}. $V_i$ represent the $i$-th layer of the variational circuit.}
    \label{fig:swap_test_circuit}
\end{figure}

The goal of the VQLS algorithm~\cite{bravo2019variational} is to produce a quantum state $\ket{\psi}=A\ket{\hat{x}}$ that
is proportional to $\ket{b}$, the representation of
$b$ in amplitudes.
The encoding of $\ket{\hat{x}}$ is performed thanks to variational circuit
$V(\mathbf{\theta})$ acting on a register that is initially prepared in the
$\ket{0}$ state. We thus have $\ket{\hat{x}}=V(\mathbf{\theta})\ket{0}$.
A classical optimization loop is then used to tune the parameters $\theta$ with respect to the minimization of a cost function.
In order to encode the problem in this cost function, we define it in the following way~\cite{bravo2019variational}:
\begin{equation}
    \mathcal{C}_\text{SWAP test}(\theta) = 1 - \abs{\bra{b}\ket{\psi(\theta)}}^2
\end{equation}
Minimizing $\mathcal{C}_\text{SWAP test}(\theta)$ will tune the parameters $\theta$ such that $\ket{\psi(\theta)}$ is close to $\ket{b}$.
The distance contribution is estimated with the circuit depicted in Figure~\ref{fig:swap_test_circuit}.

To study the robustness of the algorithm, we select a simple problem and set $A$ to the identity, and $b$ to a flat vector. Simulations with different level of noise ranging from $10^{-9}$ to $10^{-1}$ were performed by varying the number of layers and the size of the problem (number of qubits).
In order to have a statistically relevant estimation of the cost function, we run the simulations 100 times and average the final cost function obtained after convergence.

Note that the two Hadamard gates acting on the ancilla register are bias-preserving. Indeed, the first one prepares the $\ket{+}$ state and the last one measures in the $+$ basis\cite{guillaud2021error}.
\subsubsection{Results analysis}

\begin{figure*}[t]
    \centering
    \subfigure[Hardware-agnostic gate-based with Toffoli\label{fig:vqls_arb_par}]{\includegraphics[width=0.32\linewidth]{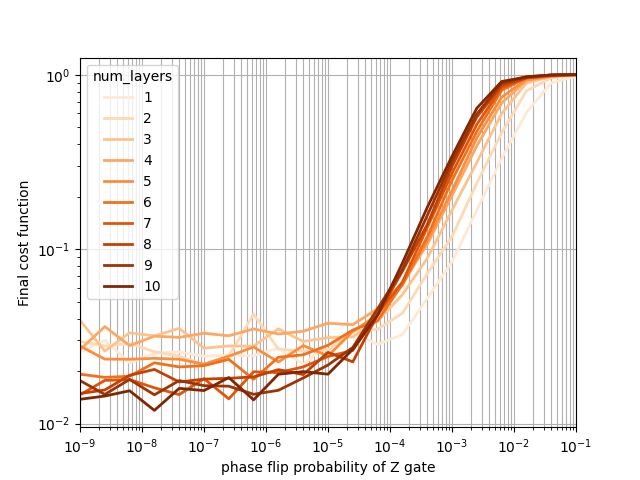}}
    \subfigure[Parallel hardware-agnostic gate-based without Toffoli\label{fig:vqls_arb_wt}]{\includegraphics[width=0.32\linewidth]{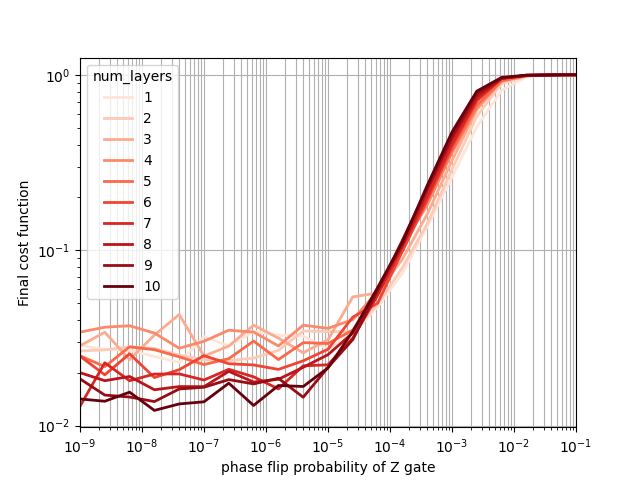}}
    \subfigure[Parallel gate-based Cat qubits\label{fig:vqls_cat_par}]{\includegraphics[width=0.32\linewidth]{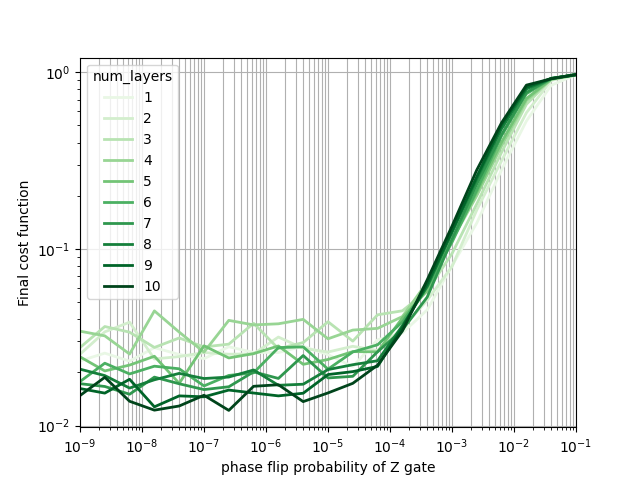}}\\
    \caption{VQLS: Evolution of the final cost function reached after convergence depending on the number of layers ranging from 1 to 10 for different noise models ~\ref{fig:vqls_arb_par} parallel hardware-agnostic with a native Toffoli. ~\ref{fig:vqls_arb_wt} parallel hardware-agnostic without a native Toffoli. ~\ref{fig:vqls_cat_par} parallel cat qubits model.}
    \label{fig:vqls_layers}
\end{figure*}
We fix the problem to 7 qubits in the main text, thus inverting matrices of size $8\times8$ (3 qubits for the target register, 3 qubits for matrix register, 1 ancilla) and only look at gate-based noise models. Appendix~\ref{sec:vqls_noise_resiliency_thresholds} regroups configurations with matrices of size $4\times4$ (5 qubits) $8\times8$ (7 qubits) $16\times16$ (9 qubits) and $32\times2$ (11 qubits).
In Figure~\ref{fig:vqls_layers}, we select the hardware-agnostic noise model with~\ref{fig:vqls_arb_par} and without~\ref{fig:vqls_arb_wt} a native Toffoli. Figure~\ref{fig:vqls_cat_par} shows the results for the cat qubits model.
For each configuration, we identify the same 3 phases of convergence depending on the level of noise introduced in the circuit.

We first observe that the noise resilient regimes reach a lower level of final cost function while increasing the number of layers.
Then, we observe that increasing the number of layers diminishes the noise resiliency threshold, which was already observed for QAOA.

\begin{figure}[H]
    \centering
    \subfigure[7 qubits]{\includegraphics[width=0.98\linewidth]{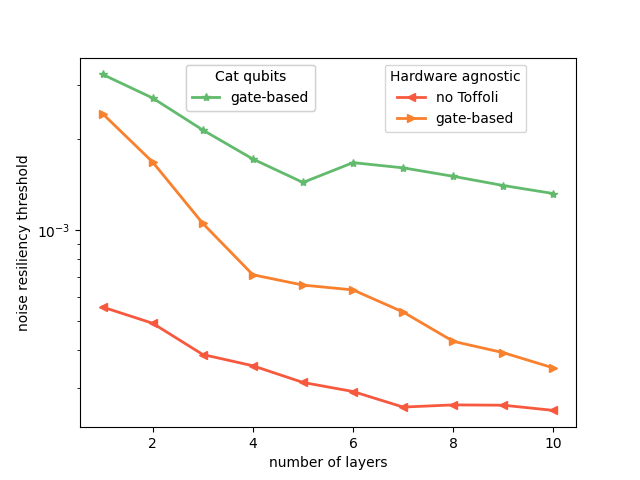}}
    \caption{VQLS noise resiliency thresholds against the number of layers for 7 qubits.}
    \label{fig:vqls_noise_resiliency_thresholds_3}
\end{figure}

In Figure~\ref{fig:vqls_noise_resiliency_thresholds_3}, we compare the behavior of the noise models.
The thresholds for the hardware-agnostic model compiled in parallel and considering that the Toffoli is a native gate are lower than the ones with the cat qubits model.

If the Toffoli gate is not considered as native and is instead compiled with other one and two qubit gates, the depth of the circuit increases and the threshold is altered towards a lower value. For a low number of layers, the main contributions to the errors originates from the Toffoli layers. As the number of layers increases, the gap between the native and non native Toffoli gates decreases.

This indicates that the fact that the Toffoli gate is native for the cat qubits model might offer an advantage over a technology that needs to compile it with other gates. In the NISQ regime, the depth introduced by the compilation of the Toffoli gate is large enough compared to the total depth of the circuit, which constrains the averaged error rate introduced in the circuit. 

\section{Conclusion}
\label{sec:conclusion}
Cat qubits are thought to be used in a fault tolerant architecture for which there exist demonstrations of reduced overhead when introducing error correcting codes compared to other technologies.
However, the intrinsic property of protection against bit-flips lead us to think that some gain could be obtained at the physical level, before introducing error correction codes.
The goal of this work was thus to quantify the convergence of Variational Quantum Algorithms on cat qubits. While finalizing our work, we became aware of a study carried out in parallel with a different approach that focuses on the performance of the QAOA algorithm with cat qubits~\cite{vikstal2023quantum}.

To reach this goal, we first started by extracting the relevant information on the noise models in the cat qubits formalism from the literature. We proposed a way to integrate the Hadamard gate, thus introducing bit-flips in the computation.

We then applied the resulting noise model to two problems of interest that can be tackled with VQAs, optimization and linear system inversion.
For optimization, we chose a problem that is frequently used for benchmarking, the MaxCut problem. VQLS was used for the system inversion on the identity with a simple target vector.
We studied the final value of the cost function after convergence for different levels of noise, for different problem sizes, and for different degrees of freedom of the algorithm - namely the number of layers at which each ansatz is repeated.
This lead us to determine thresholds for the noise levels from which the convergence can be considered in a noise-resilient regime.

A hardware-agnostic noise model allowed us to compare our hardware-specific results to a generic model, with a symmetric depolarization.
We observed that the convergence behaves similarly for the two types of models, which confirms that the hardware-agnostic model can be used to have an idea of the behavior of the algorithm in the presence of noise.
However, we observe that the acceptable regime of noise varies depending on the chosen noise model for the two algorithms studied. For QAOA, the discrepancy is low, but still increases with the number of layers. For the swap test the discrepancy significantly increases with the number of layers. By introducing compilation constraints as the presence of a native Toffoli gate, we show that the additional depth required decreases the noise resiliency threshold. Concluding on the potential advantage of the cat qubits physical model over a symmetric depolarising technology is not as straightforward as expected. One has to take into account compilation and choosing a relevant common reference between the noise models (the Z failure rate for our study).

\section*{Acknowledgement}
The authors thank Jérémie Guillaud for his support in understanding the cat qubits formalism and for giving feedback.

\bibliographystyle{unsrt}
\bibliography{main}

\begin{thebibliography}{10}

\bibitem{preskill2018quantum}
John Preskill.
\newblock Quantum computing in the nisq era and beyond.
\newblock {\em Quantum}, 2:79, 2018.

\bibitem{wang2021noise}
Samson Wang, Enrico Fontana, Marco Cerezo, Kunal Sharma, Akira Sone, Lukasz
  Cincio, and Patrick~J Coles.
\newblock Noise-induced barren plateaus in variational quantum algorithms.
\newblock {\em Nature communications}, 12(1):1--11, 2021.

\bibitem{henriet2020robustness}
Lo{\"\i}c Henriet.
\newblock Robustness to spontaneous emission of a variational quantum
  algorithm.
\newblock {\em Physical Review A}, 101(1):012335, 2020.

\bibitem{guillaud2022quantum}
J{\'e}r{\'e}mie Guillaud, Joachim Cohen, and Mazyar Mirrahimi.
\newblock Quantum computation with cat qubits.
\newblock {\em arXiv preprint arXiv:2203.03222}, 2022.

\bibitem{mirrahimi2014dynamically}
Mazyar Mirrahimi, Zaki Leghtas, Victor~V Albert, Steven Touzard, Robert~J
  Schoelkopf, Liang Jiang, and Michel~H Devoret.
\newblock Dynamically protected cat-qubits: a new paradigm for universal
  quantum computation.
\newblock {\em New Journal of Physics}, 16(4):045014, 2014.

\bibitem{guillaud2019repetition}
J{\'e}r{\'e}mie Guillaud and Mazyar Mirrahimi.
\newblock Repetition cat qubits for fault-tolerant quantum computation.
\newblock {\em Physical Review X}, 9(4):041053, 2019.

\bibitem{cohen2012processus}
Claude Cohen-Tannoudji, Jacques Dupont-Roc, and Gilbert Grynberg.
\newblock {\em Processus d'interaction entre photons et atomes}.
\newblock Edp Sciences, 2012.

\bibitem{lescanne2020engineering}
Rapha{\"e}l Lescanne.
\newblock {\em Engineering multi-photon dissipation in superconducting circuits
  for quantum error correction}.
\newblock PhD thesis, Universit{\'e} Paris sciences et lettres, 2020.

\bibitem{leghtas2015confining}
Zaki Leghtas, Steven Touzard, Ioan~M Pop, Angela Kou, Brian Vlastakis, Andrei
  Petrenko, Katrina~M Sliwa, Anirudh Narla, Shyam Shankar, Michael~J Hatridge,
  et~al.
\newblock Confining the state of light to a quantum manifold by engineered
  two-photon loss.
\newblock {\em Science}, 347(6224):853--857, 2015.

\bibitem{lescanne2020exponential}
Rapha{\"e}l Lescanne, Marius Villiers, Th{\'e}au Peronnin, Alain Sarlette,
  Matthieu Delbecq, Benjamin Huard, Takis Kontos, Mazyar Mirrahimi, and Zaki
  Leghtas.
\newblock Exponential suppression of bit-flips in a qubit encoded in an
  oscillator.
\newblock {\em Nature Physics}, 16(5):509--513, 2020.

\bibitem{guillaud2021error}
J{\'e}r{\'e}mie Guillaud and Mazyar Mirrahimi.
\newblock Error rates and resource overheads of repetition cat qubits.
\newblock {\em Physical Review A}, 103(4):042413, 2021.

\bibitem{chamberland2022}
Christopher Chamberland, Kyungjoo Noh, Patricio Arrangoiz-Arriola, Earl~T.
  Campbell, Connor~T. Hann, Joseph Iverson, Harald Putterman, Thomas~C.
  Bohdanowicz, Steven~T. Flammia, Andrew Keller, Gil Refael, John Preskill,
  Liang Jiang, Amir~H. Safavi-Naeini, Oskar Painter, and Fernando~G.S.L.
  Brand\~ao.
\newblock Building a fault-tolerant quantum computer using concatenated cat
  codes.
\newblock {\em PRX Quantum}, 3:010329, Feb 2022.

\bibitem{farhi2014quantum}
Edward Farhi, Jeffrey Goldstone, and Sam Gutmann.
\newblock A quantum approximate optimization algorithm.
\newblock {\em arXiv preprint arXiv:1411.4028}, 2014.

\bibitem{bravo2019variational}
Carlos Bravo-Prieto, Ryan LaRose, Marco Cerezo, Yigit Subasi, Lukasz Cincio,
  and Patrick~J Coles.
\newblock Variational quantum linear solver.
\newblock {\em arXiv preprint arXiv:1909.05820}, 2019.

\bibitem{schuld2020circuit}
Maria Schuld, Alex Bocharov, Krysta~M Svore, and Nathan Wiebe.
\newblock Circuit-centric quantum classifiers.
\newblock {\em Physical Review A}, 101(3):032308, 2020.

\bibitem{cerezo2021variational}
Marco Cerezo, Andrew Arrasmith, Ryan Babbush, Simon~C Benjamin, Suguru Endo,
  Keisuke Fujii, Jarrod~R McClean, Kosuke Mitarai, Xiao Yuan, Lukasz Cincio,
  et~al.
\newblock Variational quantum algorithms.
\newblock {\em Nature Reviews Physics}, 3(9):625--644, 2021.

\bibitem{qiskit_qaoa}
https://qiskit.org/textbook/ch-applications/qaoa.html.

\bibitem{vikstal2023quantum}
Pontus Vikstål, Laura García-Álvarez, Shruti Puri, and Giulia Ferrini.
\newblock Quantum approximate optimization algorithm with cat qubits, 2023.

\end{thebibliography}
\end{multicols}
\appendix
\newpage
\section{QAOA noise resiliency thresholds}
\label{sec:qaoa_noise_resiliency_thresholds}
\begin{figure*}[ht]
    \centering
    \subfigure[4 qubits]{\includegraphics[width=0.43\linewidth]{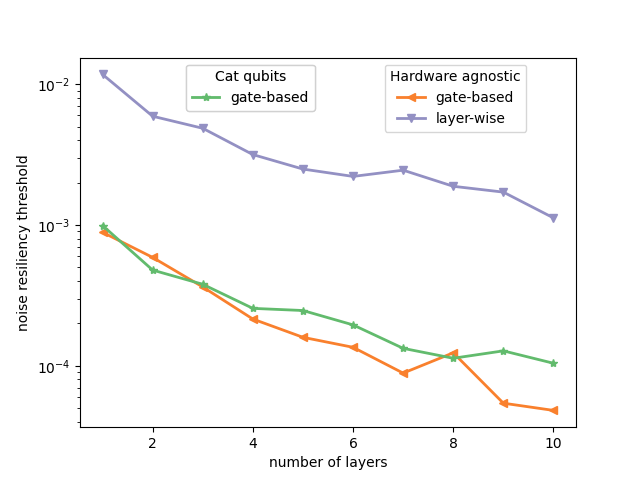}}
    \subfigure[5 qubits]{\includegraphics[width=0.43\linewidth]{qaoa_convergence_5.png}}
    \subfigure[6 qubits]{\includegraphics[width=0.43\linewidth]{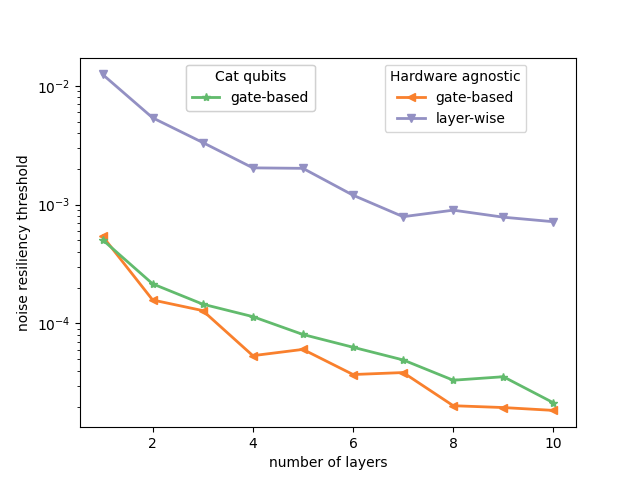}}
    \subfigure[7 qubits]{\includegraphics[width=0.43\linewidth]{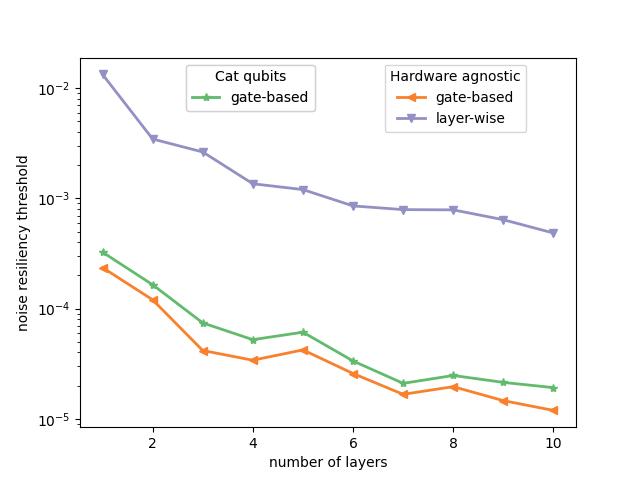}}
    \subfigure[8 qubits]{\includegraphics[width=0.43\linewidth]{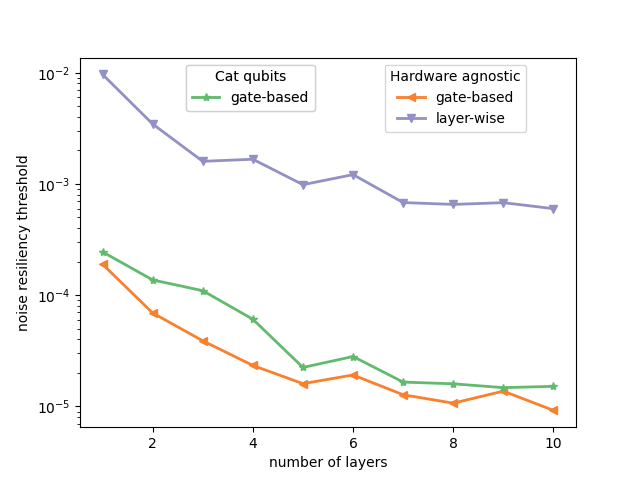}}
    \subfigure[9 qubits]{\includegraphics[width=0.43\linewidth]{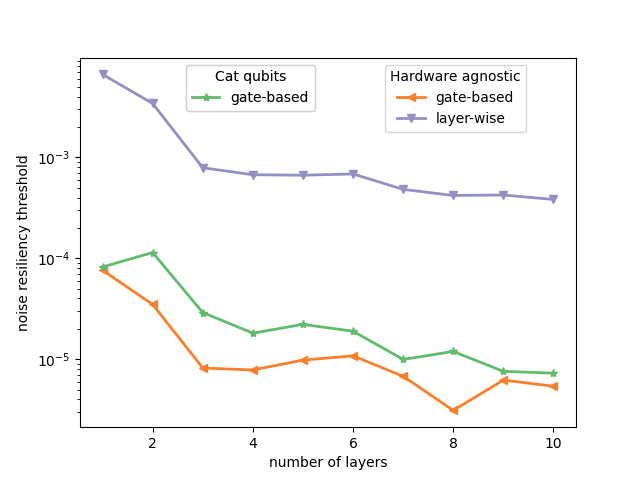}}
    \label{fig:qaoa_noise_resiliency_thresholds}
    \caption{MaxCut QAOA noise resiliency thresholds against number of layers}
\end{figure*}
\newpage
\section{VQLS noise resiliency thresholds}
\label{sec:vqls_noise_resiliency_thresholds}
\begin{figure*}[ht]
    \centering
    \subfigure[5 qubits]{\includegraphics[width=0.43\linewidth]{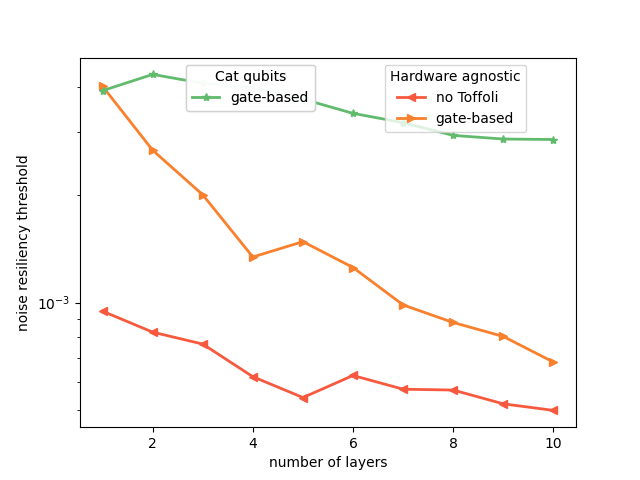}}
    \subfigure[7 qubits]{\includegraphics[width=0.43\linewidth]{swap_test_convergence_3.png}}
    \subfigure[9 qubits]{\includegraphics[width=0.43\linewidth]{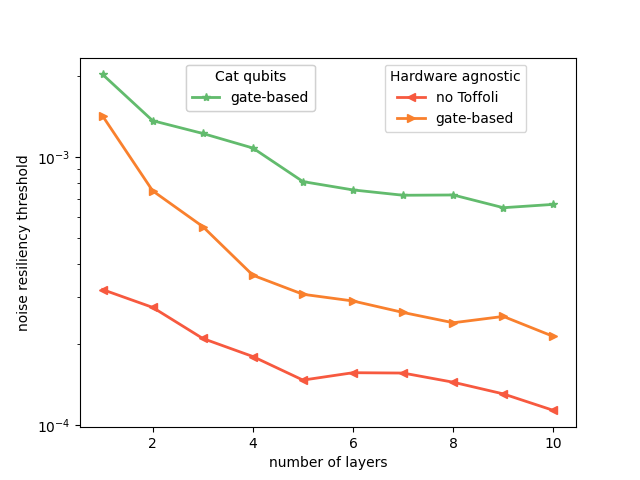}}
    \subfigure[11 qubits]{\includegraphics[width=0.43\linewidth]{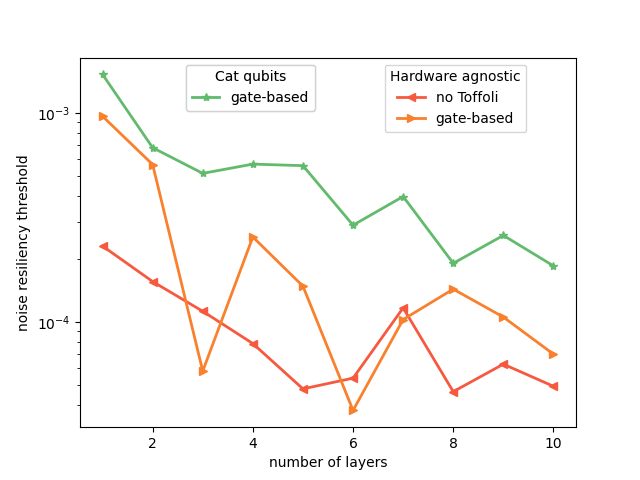}}
    \caption{VQLS noise resiliency thresholds against number of layers}
    \label{fig:vqls_noise_resiliency_thresholds}
\end{figure*}

\end{document}